# Extreme ultraviolet emission from dense plasmas generated with sub-10-fs laser pulses

J. Osterholz[1,*], F. Brandl[1], M. Cerchez[1], T. Fischer[1], D. Hemmers[1], B. Hidding[1], A. Pipahl[1], G. Pretzler[1], S.J. Rose[2], and O. Willi[1]

[1] Institute for Laser and Plasmaphysics, Heinrich-Heine-University, Düsseldorf, Germany

[2] Blackett Laboratory, Imperial College, London, United Kingdom

**Abstract**

The extreme ultraviolet (XUV) emission from dense plasmas generated with sub-10-fs laser pulses with varying peak intensities up to $3*10^{16}$ W/cm$^2$ is investigated for different target materials. *K* shell spectra are obtained from low *Z* targets (carbon and boron nitride). In the spectra a series limit for the hydrogen and helium like resonance lines is observed indicating that the plasma is at high density and pressure ionization has removed the higher levels. In addition, *L* shell spectra from titanium targets were obtained. Basic features of the *K* and *L* shell spectra are reproduced with computer simulations. The calculations include hydrodynamic simulation of the plasma expansion and collisional radiative calculations of the XUV emission.

[*] Corresponding author, e-mail: jens.osterholz@uni-duesseldorf.de

**I. Introduction**

The interaction of solid targets with intense sub-picosecond laser pulses has been studied in many laboratories in the past. The absorption of laser energy on a short time scale leads to the generation of dense plasmas with temperatures typically between 10 and 1000 eV [1-9]. Matter under this condition can be found for example in the interior of stars [10, 11], Jovian planets [12] and in inertial confinement fusion (ICF) pellets [13]. Dense plasmas are an intense source of x-ray and XUV radiation. At high density the energy levels of bound electrons can be significantly affected due to the interaction with surrounding particles. Several effects as e.g. plasma polarization shift, satellite emission, pressure ionization, continuum lowering or the Inglis Teller effect have been observed [1-9]. These effects have an important impact on the opacity and the radiative transport in the plasma. Therefore spectroscopic studies of laser generated plasmas are an important tool in the context of laboratory astrophysics and ICF. Typically, $K$ shell spectra are used to derive the plasma conditions. Due to their more complicated structure $L$ shell spectra are less suited for plasma diagnostics, however they play an important role for the energy transport in dense plasmas. The analysis of the x-ray and XUV emission is often complicated by the fact that laser generated plasmas are highly transient with strong gradients of the spatial density and temperature profiles. For an accurate calculation of the x-ray spectra, the temporal and spatial evolution of the plasma conditions has to be taken into account.

For the production of dense plasmas the energy transfer from the laser to the target and the hydrodynamic expansion of the plasma during the interaction play important roles [14-17]. The maximum density which can be achieved is limited by the laser pulse contrast and duration. Typically, the laser cannot penetrate into the solid region of the target as it is reflected at the critical surface where the density is much smaller than solid. A fraction of the laser energy is transferred to the electrons e.g. by resonance absorption or $\vec{j} \times \vec{B}$ heating and is then transported to the dense region of the target by the electrons. Consequently, a small plasma scale length during the interaction is favourable for an efficient energy transfer to the dense region of the target. In order to reduce the expansion of the plasma during the interaction, tamped targets have been used in some experiments [1,2].

Different mechanisms of the energy transfer are possible when the plasma scale length is much smaller than the laser wavelength during the interaction. Under these conditions, a large fraction of the laser energy can be transferred directly to the solid region of the target. Absorption coefficients of about 80% have been reported in [18]. This new regime of laser plasma interaction requires laser systems with an excellent contrast and extremely short pulses.

In this work close to solid density plasmas are produced using laser pulses with an excellent contrast and less than 10 fs duration with peak intensities up to $3*10^{16}$ W/cm$^2$. The XUV emission from the plasma is measured with a spectrograph covering the wavelength range between 1 and 8 nm. *K* shell spectra from BN and C targets as well as *L* shell spectra from Ti targets are obtained. The *K* shell spectra basically consist of the Ly$_\alpha$ and He$_\alpha$ lines, whereas the higher series lines are suppressed. This series limit indicates that the plasma is at high density. The measured spectra are in good agreement with computer simulations combining hydrodynamic calculations of the plasma expansion with collisional radiative calculations of the XUV emission of the plasma. Thus, the model takes into account the highly transient nature and the temperature and density profiles of the plasma.

## II. Experimental setup

The experiments were carried out with a CPA laser system similar to that described in [19]. The pulses from a titanium:sapphire oscillator are stretched using a SF57 glass block and an acousto-optical dispersive filter and are amplified with a titanium:sapphire multipass amplifier. After compression in a prism compressor, the pulse energy was 700 µJ with 30 fs duration at 1 kHz repetition rate. Further compression was achieved in a second compressor as described in [20]. In this compressor the pulses were transmitted through a noble gas filled fused silica hollow fiber with 1 m length and 200 µm diameter. Due to self-phase modulation the spectral bandwidth of the laser pulses increased from 30 nm to 120 nm. The pulses were then compressed with a set of eight specially designed chirped mirrors to a pulse length of 8 fs which was measured with a second-order autocorrelator. The pulse contrast was experimentally determined using a high dynamic range third-order autocorrelator. The diagnosis reveals a contrast ratio of $10^5$ for times larger than 1 ps before the main pulse and better than $10^8$ for the amplified spontaneous emission prepulse. Due to transmission losses in the second compressor and the mirrors in the beamline, the intensity on target was reduced to 200 µJ. The laser pulses were focused with an f/3 off-axis parabola. The focal spot with a diameter of 10 µm contained 50% of the laser energy. The peak intensity is $3 \times 10^{16}$ W/cm$^2$ corresponding to an average intensity of $10^{16}$ W/cm$^2$ in the focal region.

For the plasma diagnostics spectroscopy of the *K* shell resonance lines is often used. For the laser parameters of this experiment it is expected that the hydrogen-like and helium-like ionization stages required for the *K* shell emission can only be generated in targets with a small atomic number. Therefore carbon and boron nitride targets were used in the experiments. The *K* shell emission from these low Z elements is in the XUV spectral range. In addition, this spectral range is suited for spectroscopic studies of the *L* shell emission of higher Z targets as e.g. titanium.

An XUV spectrograph with high detection efficiency similar to that described in [21] was constructed. A drawing of the spectrograph is shown in Fig. 1. The spectrograph was equipped with varied line-spacing flat-field grazing-incidence gratings with a nominal groove spacing of 1200 or 2400 lines/mm [22]. A modified Kirkpatrick-Baez arrangement [23] consisting of two gold coated, spherical mirrors is used for collecting the XUV emission from the plasma. The first mirror (radius of curvature 7 m, grazing angle 2.2°, diameter 50.8 mm) produces a line focus of the XUV light at the virtual spectrometer entrance slit, whereas the second mirror (radius of curvature 10 m, grazing angle 2.6°, diameter 50.8 mm) images the plasma onto the image plane of the detector in one dimension. The spectra were acquired with an intensified CCD camera. The pixel size of the CCD chip is 14 μm x 14 μm with a total of 2048 x 2048 pixels. The image intensifier consists of a phosphor scintillator coated with a 100 nm Al layer and a microchannel plate. In the experiment a spectral resolution of $\lambda/\Delta\lambda \approx 100$ at 5 nm was achieved. The nominal inverse linear dispersion of the spectrograph is 0.6 nm/mm or 0.3 nm/mm for the 1200 or 2400 lines/mm grating, respectively.

The p-polarized laser was incident onto the target at an angle of 45°. Depending on the energy of the XUV emission, the spectra were time integrated over a time interval of up to 800 seconds at the 1 kHz repetition rate of the laser, corresponding to a maximum total number of $8 \times 10^5$ laser shots. During the acquisition the target was moved laterally in order to keep the target surface at the focus position.

**III. Results and Discussion**

A typical *K* shell spectrum obtained from a carbon target with the sub-10-fs laser pulses is shown in Fig. 2. The spectrum consists only of the C VI 1*s*-2*p* (C Ly$_\alpha$, 3.37 nm) and the C V 1*s*$^2$-1*s*2*p* (C He$_\alpha$, 4.03 nm) lines. There may be a small contribution of the C V 1*s*$^2$-1*s*3*p* (C He$_\beta$, 3.50 nm) in the spectrum. However, this line is only slightly above the noise level. In contrast spectra generated with longer laser pulses typically contain the higher series lines. This is demonstrated by the second carbon spectrum in Fig. 2 which was obtained using a frequency doubled Nd:YAG laser with 8 ns, 120 mJ laser pulses at an intensity of $5 \times 10^{12}$ W/cm$^2$ in the same experimental setup. In the spectrum generated with the longer laser pulses the hydrogen and helium like series including a recombination continuum are clearly visible. It is evident that the higher series lines are not emitted for the sub-10-fs pulses.

In a second experiment the *K* shell emission from a BN target irradiated with the sub-10-fs laser pulses was investigated. In the spectrum in Fig. 3 the N VII 1$s$-2$p$ (N Ly$_\alpha$, 2.48 nm), N VI 1$s^2$-1$s$2$p$ (N He$_\alpha$, 2.88 nm), B V 1$s$-2$p$ (B Ly$_\alpha$, 4.86 nm), B IV 1$s^2$-1$s$2$p$ (He$_\alpha$, 6.03 nm) and the B IV 1$s^2$-1$s$3$p$ (B He$_\beta$, 5.27 nm) are labelled. Similar to the C spectrum the higher series lines are not observed.

The lack of the higher series lines in the *K* shell spectra obtained with sub-10-fs laser pulses is an indication for pressure ionization. Pressure ionization is a consequence of the overlap of the electron wavefunctions in the higher states of adjacent ions in dense plasmas. Under these conditions the electrons in those states are essentially free and the higher levels do not exist anymore. Pressure ionization effectively reduces the ionization potential of the ions. According to [24] the ionization potential depression $\Delta E$ (in eV) is:

$$\Delta E [eV] = 2.16 \cdot 10^{-7} \frac{z}{r_i [cm]} \left\{ \left[ 1 + \left(\frac{\lambda_d}{r_i}\right)^3 \right]^{2/3} - \left(\frac{\lambda_d}{r_i}\right)^2 \right\} \quad (1)$$

where z is the ion charge, $\lambda_d = (\varepsilon_0 kT_e/n_e e^2)^{1/2}$ is the Debye radius, $\varepsilon_0$ is the vacuum dielectric permittivity, k is the Boltzmann constant, $T_e$ is the electron temperature, $n_e$ is the electron density, e is the electron charge, $r_i = (4\pi n_i/3)^{-1/3}$ is the ion sphere radius, and $n_i$ is the ion density.

The plasma density can be estimated from the observed series limit using Equation 1. To this end we plotted the ionization potential of the hydrogen and helium like boron versus the plasma density for three different temperatures in Fig. 4. The calculation was carried out assuming an average ion charge of 4.5 in agreement with the observed XUV lines. For the BN plasma the density range which suppresses the B Ly$_\beta$ but allows for the emission of the B He$_\alpha$ is between 0.36 and 1.37 g/cm$^3$ at a temperature of 100 eV. For the carbon plasma at 100 eV the suppression of the C He$_\beta$ requires a density of more than 0.55 g/cm$^3$.

A more precise calculation of pressure ionization under steady state conditions was carried out with the IMP code [25]. IMP calculates the potential around an ion in the plasma using the high-temperature Thomas-Fermi model. For this potential the Schrödinger equation is solved for different orbitals. In this way we determine which orbitals are and which orbitals are not allowed in the plasma. For a temperature of 100 eV the range of densities which allow for the emission of the Ly$_\alpha$ and suppress the Ly$_\beta$ lines is between $\rho_{min}$ = 0.3 g/cm$^3$ and $\rho_{max}$ = 0.8 g/cm$^3$ for BN and between $\rho_{min}$ = 0.2 g/cm$^3$ and $\rho_{max}$ = 1.3 g/cm$^3$ for C.

Although steady state calculations are useful for an estimation of the plasma conditions, they are not sufficient for the detailed understanding of the measured spectra. The quantitative analysis of the XUV emission of laser produced plasmas is often complicated by the high gradients of the temperature and density and the rapid temporal evolution of the plasma parameters. Time resolved measurements with x-ray streak cameras with sub ps resolution have contributed to a more detailed understanding of the x-ray emission from highly transient, expanding plasmas [6, 7]. However, this technique does not apply here because under our experimental conditions the emitted XUV intensity is low and the duration of the XUV emission is extremely short, as discussed below.

For a detailed quantitative analysis of the time integrated XUV spectra, simulations of the plasma expansion including the strong temperature and density gradients, and the highly transient nature of the plasma are required. Absorption experiments with solid, conducting targets have recently shown that the energy transfer from the sub-10-fs laser pulses to the target can be well described by particle in cell (PIC) simulations [18]. The measurements were carried out under identical experimental conditions using the same laser system as for the XUV experiment. A good agreement between the measured absorption and the PIC simulations was observed for a preplasma scale length L in the order of $L/\lambda \sim 1\%$. The small size of the preplasma is attributed to the short pulse duration and the excellent contrast of the laser.

Although the PIC code is ideal for a detailed understanding of the absorption processes, the calculation of the XUV emission requires the simulation of the plasma expansion over much larger temporal and spatial scales. Because this is still a challenging task for PIC codes, we used the 1D Lagrangian hydrocode MULTI-fs [26] for the analysis of the XUV spectra, instead. MULTI-fs has been successfully used to calculate the interaction of ultra-short intense laser pulses with dense plasmas [1, 2, 9]. It is noted that the hydrocode does not explicitly take into account kinetic effects, but a good agreement between the absorption and the XUV spectra measured in the experiment could be obtained by adjusting the values for the preplasma scale length and the flux limiter. These results are in agreement with observations in [27] and indicate that the hydrodynamic simulations can calculate the expansion of the dense plasma produced with sub-10-fs laser pulses.

For the MULTI-fs calculations in this work we implemented a prepulse generating a preplasma with a scale length of a few nanometers, similar to the scale length used in the PIC simulations. The simulations were carried out for a series of times $t_j$ with a time interval $\Delta t_j = t_{j+1} - t_j$ varying from 0.1 fs close to the laser peak to longer intervals at the later stage of the expansion. The target was divided into a total number of N=160 Lagrangian layers as shown in Fig. 5. A small value of 0.05 nm was chosen for the initial nominal layer thicknesses at the target front side to resolve the high temperature and density gradients. In the following $x_{kj}$ denotes the position of the $k^{th}$ Lagrangian layer at time $t_j$. For the laser pulse a Gaussian shape with a FWHM of 8 fs was chosen. The equation of state (EOS) of the target material was obtained from the SESAME database. Simulations were carried out for different values of the flux limiter f. The best agreement with the experimental data was obtained for small values of f. This might be attributed to the small focus spot size in the experiment resulting in 2D effects. In the following calculations a value of f=0.001 was used.

For the simulation of the measured spectra, we used the plasma parameters obtained from the MULTI-fs code as input for FLY simulations. The FLY code suite has been widely used for the calculation of the atomic kinetics and the XUV emission from laser generated plasmas [28]. From the FLY simulations the temporal evolution of the opacities $\kappa(t_j, x_{k,j})$ and emissivities $\varepsilon_\nu(t_j, x_{k,j})$ for each individual layer is obtained. The simulations include time dependent population dynamics and detailed line shape calculations of the emitted resonance lines.

The specific intensity $I_\nu$ emitted from the plasma was obtained for each time step by solving the radiative transport equation

$$\frac{dI_\nu}{d\tau} = -\kappa \cdot I_\nu + \varepsilon_\nu \quad (2)$$

where $d\tau = \kappa \cdot dx$ is the optical depth of a layer with thickness dx. Equation (2) was integrated numerically by adding up the contributions of each layer:

$$I_\nu(t_j, x_{k+1,j}) = S_\nu(t_j, x_{k+1,j}) + \exp(-\tau_{k+1,j}) \cdot (I_\nu(t_j, x_{k,j}) - S_\nu(t_j, x_{k+1,j})) \quad (3)$$

where $S_\nu = \varepsilon_\nu / \kappa$ is the source function and $\tau_{k+1,j} = \kappa(t_j, x_{k+1,j})(x_{k+1,j} - x_{k,j})$ is the optical thickness of the layer number k+1. Finally, the time integrated spectrum D(ν) was obtained by adding up the contributions to the specific intensity for each time step:

$$D(\nu) = \sum I_\nu(t_j, x_{N,j}) \Delta t_j \quad (4)$$

with the time step size $\Delta t_j = t_j - t_{j-1}$.

The plasma conditions for the C plasma calculated with MULTI-fs for three different times are plotted in Fig. 6. The time integrated XUV spectrum calculated as described above is shown in Fig. 7. In the simulated spectrum only small contributions of the higher series lines are observed. The ratios of the Ly$_\beta$ / Ly$_\alpha$ and He$_\beta$ / He$_\alpha$ are smaller than 0.09 and 0.06, respectively. This is close to the noise level of the experiment, and consequently the lines are not observed in the measured spectra. In the hydrodynamic simulations there is only a small preplasma scale length of a few nm at t=-10fs right before the laser pulse. Due to the small preplasma scale length, there is only a weak contribution of the low density part of the plasma to the overall XUV emission. The emission is dominated by the region of the plasma where the density is high and the higher series lines are suppressed by pressure ionization. At later times the plasma scale length becomes larger, but the plasma cools down quickly. This is demonstrated in Fig. 8 where the temperature and density history of the Lagrangian layer number 92 with an initial position of x=2.7nm are plotted. Close to the laser peak the temperature rapidly increases to about 200 eV. When the temperature is at its maximum, the density is still larger than 1 g/cm$^3$. The inset in Fig. 8 shows the temporal evolution of the intensities of the C Ly$_\alpha$ and C He$_\alpha$ lines for layer number 92. The line intensities drop off with a time constant on a ps scale. Consequently the emission of the hydrogen and helium-like resonance lines is dominated from the period where the plasma is at high density. In the simulations it was found that the thickness of the emitting plasma region is very small. For example at the time 10 fs after the laser peak, the C Ly$_\alpha$ line is emitted from a region with a thickness of about 10 nm (FWHM). This value is in agreement with the expansion model in [29] where the initial thickness of the emitting plasma layer is equal to the electron skin depth. Similar results were obtained for the BN plasma. In Fig. 9 the calculated ratios of the Ly$_\alpha$ / He$_\alpha$ lines are shown for B, C and N. The calculated values agree within the experimental error of 20% with the measured intensity ratios, indicating that the computer models used in this work are well suited for the calculation of XUV spectra generated with sub-10-fs laser pulses.

The ratio of the Ly$_\alpha$ to the He$_\alpha$ lines is extremely sensitive to the plasma temperature and thus to the incident laser intensity. Therefore the intensity ratio of these lines is often used to derive the plasma temperature. We measured the XUV emission from BN plasma for different values of the laser intensity. The laser intensity was varied by moving the target along the laser propagation direction (z-axis in Fig. 1). The measured ratios of the B Ly$_\alpha$ /B He$_\alpha$ lines are shown in Fig. 10. For the intensity scaling of the plot in Fig. 10 the laser beam was imaged at different z-positions with a CCD camera, yielding a Rayleigh length of $z_R$=30µm. In the experiment the ratio of the B Ly$_\alpha$ / He$_\alpha$ line decreases approximately linear from 1.5 to 0.8 for laser intensities between 3 and 0.5 *10$^{16}$ W/cm$^2$. For smaller intensities the ratio decreases quickly. For a quantitative analysis the XUV emission from the plasma was calculated for varying intensity using the MULTI-fs and FLY codes as described above. The result of the calculation is also shown in Fig. 10. There is a good qualitative agreement between the measured and the calculated line intensity ratios. Deviations between experiment and theory might be due to the error in the intensity scaling for different target z positions. The absence of the higher series lines for low Z materials is a remarkable result of the experiments. It is interesting to investigate the XUV emission in higher Z materials in the same spectral range. Due to the higher nuclear charge, the size of the inner shells is smaller and it is expected that higher series lines are emitted. For Ti XIII, for example, Equation 1 predicts an ionization potential depression of about $\Delta E=72$ eV at 0.25 times solid density and a temperature of 200eV. This value is much smaller than the ionization potential of 788 eV and does not lead to pressure ionization of the *M* shell electrons.

An XUV spectrum obtained from Ti plasma is shown in Fig. 11. Due to the larger number of electrons and the more complicated electron configuration, the *L* shell spectra contain a larger number of spectral lines. For the identification of the lines observed in the spectrum, computer simulations were carried out. Because the FLY code is limited to ions with a maximum number of three electrons, a different approach based on the FLYCHK code [30,31] was used to analyze the data. FLYCHK uses a set of superconfigurations to calculate the charge states and population distributions for ions with more than three electrons. For the time dependent calculation of the plasma conditions, an expansion model described in [29] was used. In the model it is assumed that a fraction of the laser energy is converted to thermal energy within a layer with an initial thickness corresponding to the electron skin depth. The plasma expansion is calculated assuming an adiabatic equation of state. The calculated expansion history is then used as input for the FLYCHK calculations. The result of the simulation is also shown in Fig. 11. According to the simulations the XUV emission in the investigated spectral range is basically from *L* shell excited states to the ground state. In Fig. 11 the lines from inner shell transitions from Ti XII to Ti XVII are labelled. The labels are explained in Table I. Basic features of the experimental data are reproduced in the synthetic spectrum. The main difference from the measured spectrum is the smaller intensity in the wavelength range around 2.74 nm. This might be attributed to the expansion model which considers only a single plasma layer. More precise results might be obtained from hydrodynamic simulations of the plasma expansion in combination with calculations of the *L* shell emission. This task is planned for the future.

Although laser generated plasmas are highly transient, conditions close to LTE can be approached quickly at high density [9]. This has been confirmed in FLY simulations where LTE conditions were reached a few 100 fs after the laser peak. It is also interesting to calculate the ion-ion coupling parameter $\Gamma = Z^2 e^2 / 4\pi\varepsilon_0 r_i kT$ for the plasma conditions of our experiment. Here Z is the ion charge and $r_i = (3/4\pi n_i)^{1/3}$ is the ion sphere radius. For $\Gamma > 1$ the Coulomb energy of the ions exceeds their kinetic energy and the plasma is strongly coupled. In the simulations the ion-ion coupling parameter is close to 1 for the BN and C plasmas, and significantly larger than 1 for the Ti plasma within an interval of more than 10 picoseconds after the laser peak.

## IV. Summary and conclusion

Dense plasmas were produced by irradiating solid targets with sub-10-fs, high contrast laser pulses. The absence of the higher series lines in the *K* shell spectra obtained from BN and C targets is explained by pressure ionization. Computer simulations including hydrodynamic expansion of the plasma, collisional radiative calculations of the opacities and emissivities and radiative transport in the plasma are in good agreement with the spectra measured for different laser intensities. In addition, *L* shell spectra from Ti plasma were generated with the sub-10-fs laser pulses. Basic features of the spectra are reproduced by computer simulations. The results demonstrate that an efficient energy transfer from the laser pulses to the dense region of the target can be achieved with the laser system used for this experiment. The emission is from a narrow region in the target with dimensions in the order of 10 nm, much smaller than in experiments with longer laser pulses. We conclude that high-contrast, few-cycle laser pulses are an interesting tool for spectroscopic studies of dense plasmas.

**Acknowledgements**

The authors thank K. Eidmann for valuable discussions about MULTI-fs. This work has been performed within the SFB/Transregio TR 18 and GRK 1203.

**Tables**

| Label | Transition | Wavelength range [nm] |
|---|---|---|
| 1 | Ti XII $1s^22s^22p^63l$-$1s^22s^22p^53l'$ | 2.66-2.80 |
| 2 | Ti XIII $1s^22s^22p^6$-$1s^22s^22p^55d$ | 1.77-1.79 |
| 3 | Ti XIII $1s^22s^22p^6$-$1s^22s^22p^54l$ | 1.92-2.01 |
| 4 | Ti XIII $1s^22s^22p^6$-$1s^22s^22p^53p$ | 2.10-2.11 |
| 5 | Ti XIII $1s^22s^22p^6$-$1s^22s^22p^53d$ | 2.34-2.40 |
| 6 | Ti XIII $1s^22s^22p^6$-$1s^22s^22p^53s$ | 2,66-2.70 |
| 7 | Ti XIV $1s^22s^22p^5$-$1s^22s^22p^44d$ | 1.76-1.79 |
| 8 | Ti XIV $1s^22s^22p^5$-$1s^22s^22p^43d$ | 2.13-2.25 |
| 9 | Ti XIV $1s^22s^22p^5$-$1s^22s^22p^43s$ | 2.37-2.53 |
| 10 | Ti XV $1s^22s^22p^4$-$1s^22s^22p^33d$ | 2.01-2.11 |
| 11 | Ti XV $1s^22s^22p^4$-$1s^22s^22p^33s$ | 2.21-2.34 |
| 12 | Ti XVI $1s^22s^22p^3$-$1s^22s^22p^23d$ | 1.90-2.01 |
| 13 | Ti XVII $1s^22s^22p^2$-$1s^22s^22p3d$ | 1.80-1.88 |
| 14 | Ti XVII $1s^22s^22p^2$-$1s^22s^22p3s$ | 1.94-1.97 |

Table I. List of transitions from *L* shell excited levels contributing to the XUV emission of the titanium plasma.

**Figures**

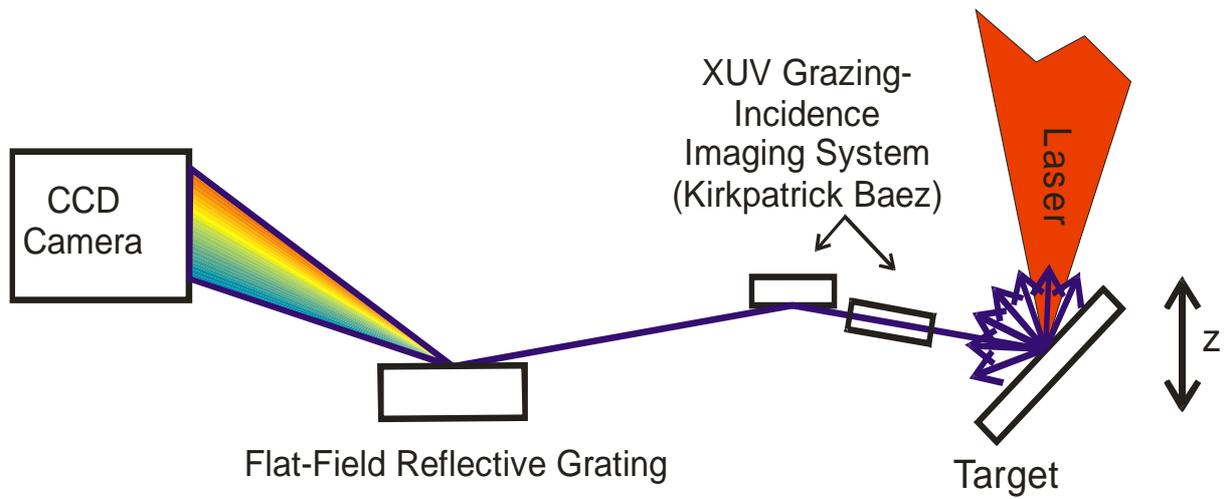

Fig. 1. Experimental setup. The emission from the plasma was analyzed with an XUV spectrograph.

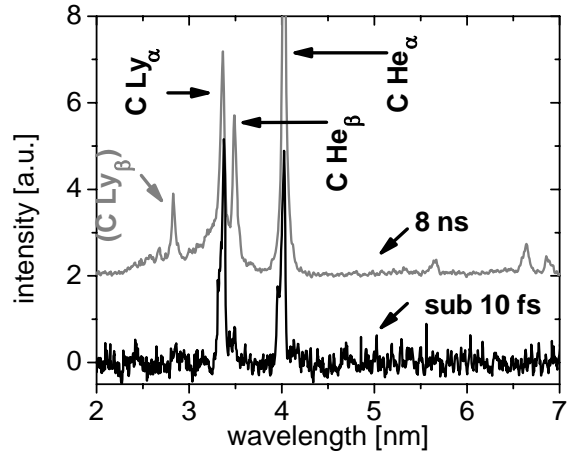

Fig. 2. XUV spectra obtained from a carbon target. Black line: sub-10-fs laser pulses, grey line: 8 ns laser pulses, plotted with an offset. For the sub-10-fs laser pulses only the C Ly$_\alpha$ and C He$_\alpha$ lines are observed, whereas pressure ionization suppresses the higher series lines. For the longer laser pulses the whole hydrogen and helium like series including a recombination continuum are observed.

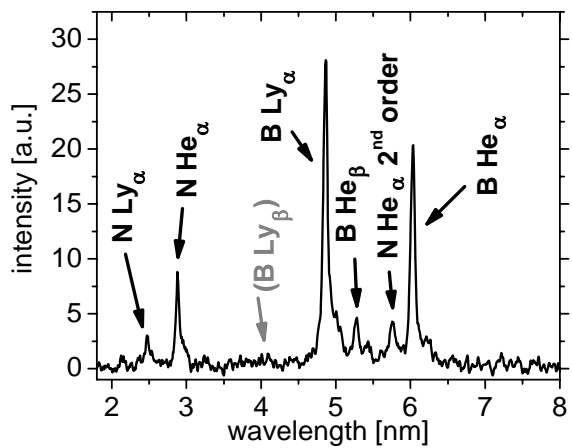

Fig. 3. XUV spectra obtained from a boron nitride plasma generated by sub-10-fs laser pulses. Due to pressure ionization the higher series lines including the B Ly$_\beta$ are not observed in the spectrum.

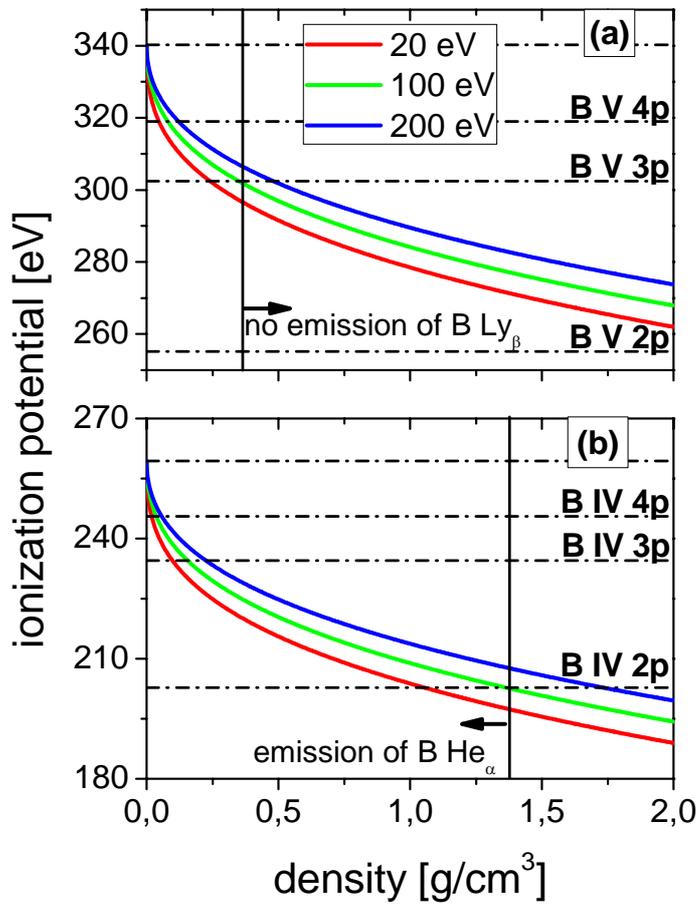

Fig. 4. Solid lines: Ionization potential of hydrogen-like (a) and helium-like (b) boron in a boron nitride plasma versus density for three different temperatures. Dash-dotted lines: Excited state energy levels. At high density the higher levels do not exist anymore due to ionization potential depression. The density ranges which allow for the emission of the B He$_\alpha$ but not the B Ly$_\beta$ lines are indicated for a temperature of 100 eV.

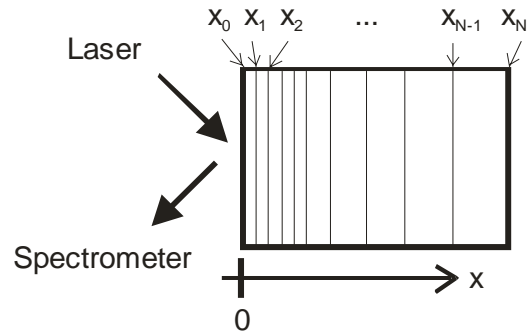

Fig. 5. Geometry for the simulation of the plasma expansion. The boundaries of the Lagrangian layers in the hydrodynamic simulations are denoted by $x_i$. The initial position of the target surface is at x=0.

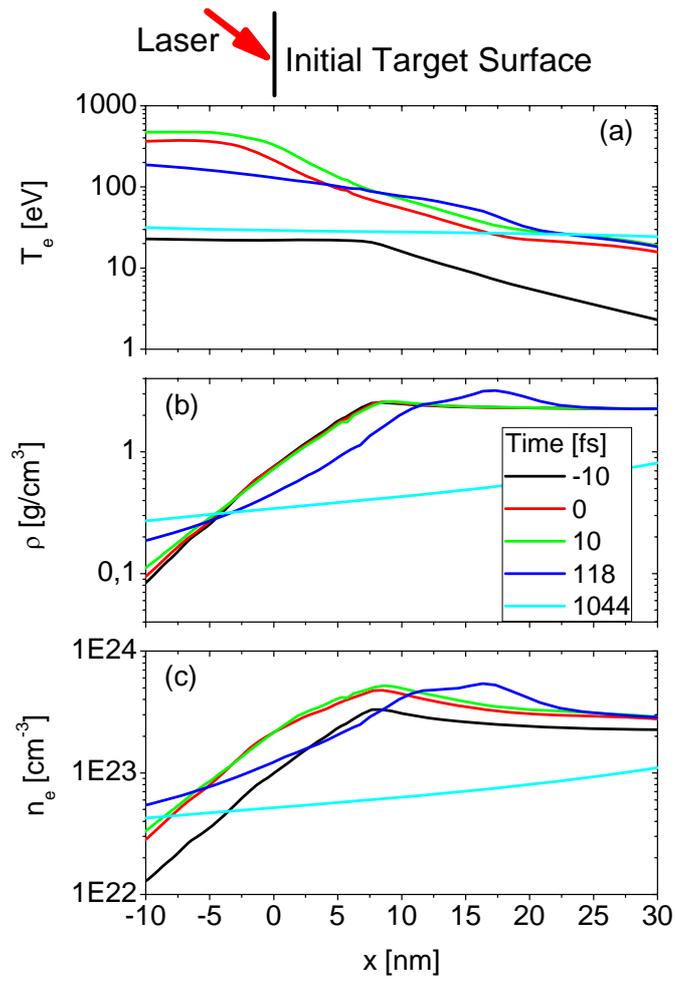

Fig. 6. Calculation of electron temperature (a), mass density (b) and electron density (c) in carbon plasma obtained with a 1 dimensional Lagrangian hydrocode for different times. The time t=0 corresponds to the laser peak. The initial position of the target surface is at x=0. The laser pulse with a peak intensity of $3*10^{16}$ W/cm$^2$ and a duration of 8 fs is incident from the left.

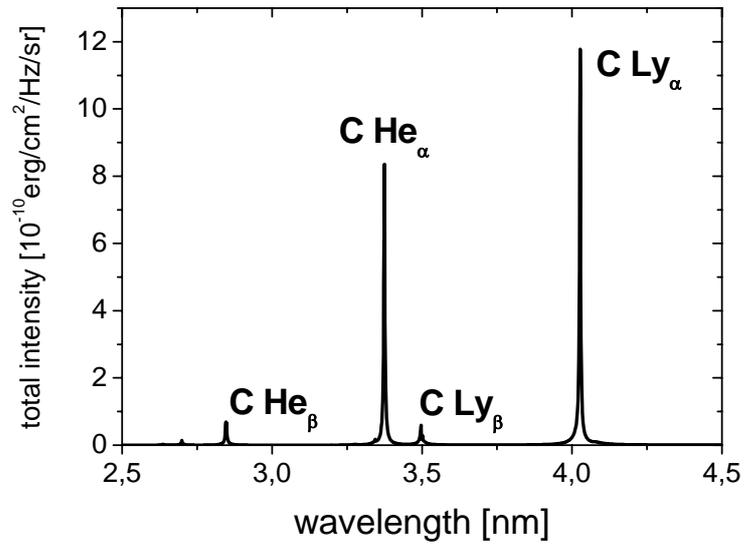

Fig. 7. Computer simulation (MULTI-fs and FLY) of the XUV emission from carbon plasma. The $Ly_\alpha$ and $He_\alpha$ lines are clearly visible. The contributions of the higher series lines are strongly suppressed by pressure ionization.

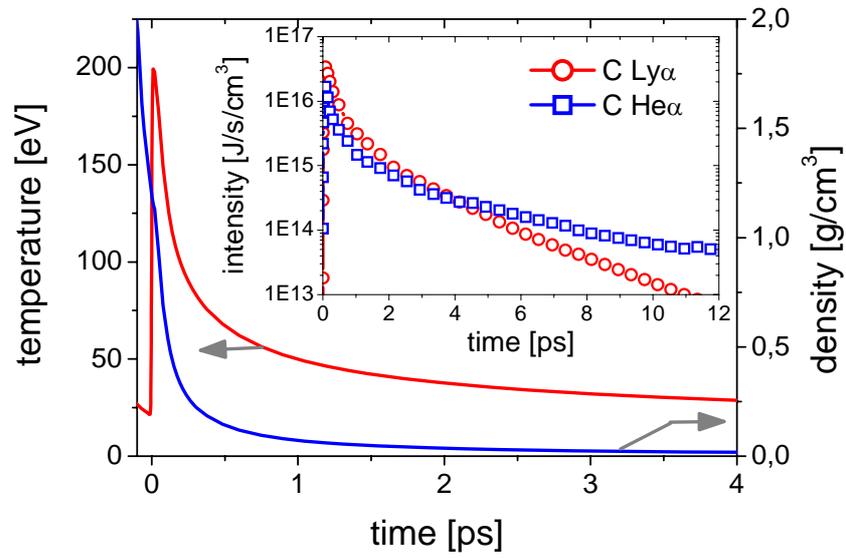

Fig.8. Temporal evolution of the electron temperature and mass density of the Lagrangian layer number 92. The peak electron temperature is 200 eV. The inset shows the temporal evolution of the intensities of the C Ly$_\alpha$ (red circles) and C He$_\alpha$ (blue boxes) lines emitted by the layer.

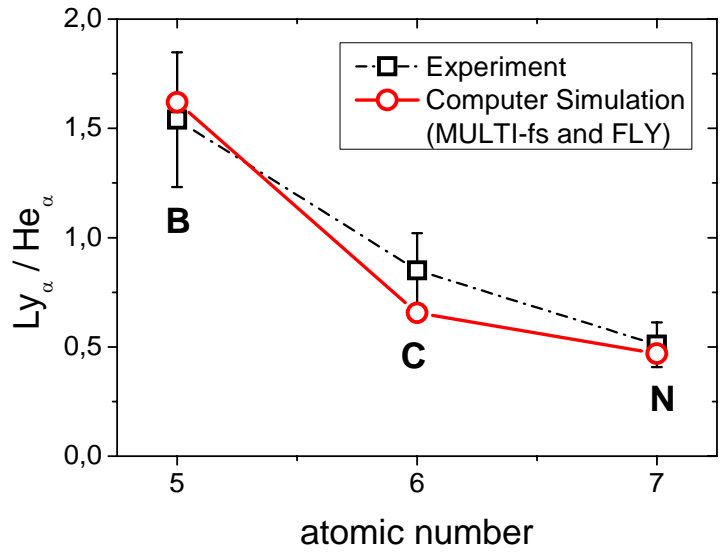

Fig. 9. Intensity ratios of the boron, carbon and nitrogen $Ly_\alpha / He_\alpha$ resonance lines observed in the experiment (boxes) and obtained from computer simulations (circles).

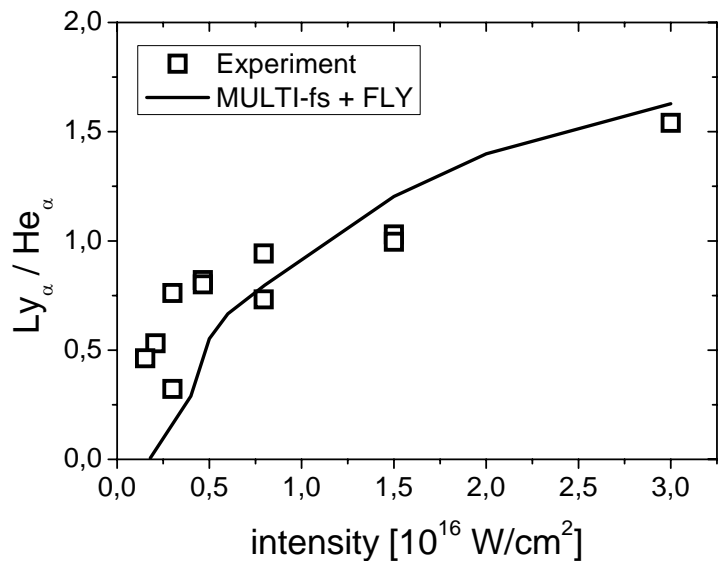

Fig. 10. Intensity ratios of the boron $Ly_\alpha / He_\alpha$ resonance lines versus laser intensity. Boxes: experiment, line: computer simulation (MULTI-fs and FLY)

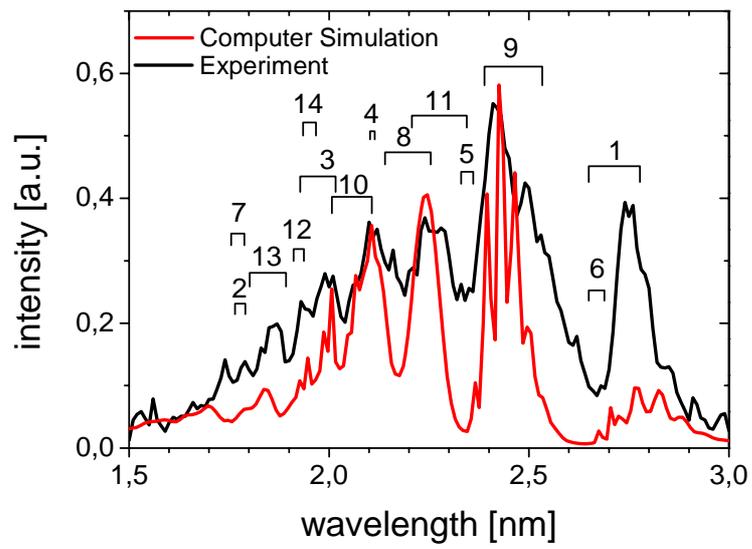

Fig. 11. *L* shell spectrum obtained from a Ti target irradiated with the sub-10-fs laser pulses. Black line: experiment, red line: computer simulation. The labels of the transition are explained in Table I.